\documentclass[final,5p,times,twocolumn]{elsarticle}

\usepackage{hyperref}

\journal{Phys.\ Lett.\ B}

\usepackage{graphicx}
\usepackage{amssymb}
\usepackage{amsmath}
\usepackage{amsthm}
\usepackage[normalem]{ulem}
\biboptions{sort&compress}
\usepackage{slashed}
\usepackage{color,rotating}
\usepackage{bbm}
\usepackage{array}
\usepackage[utf8]{inputenc}
\usepackage{subcaption}
\usepackage{multirow}
\usepackage{xcolor}
\usepackage{multirow}

\allowdisplaybreaks

	\newcommand{\p}{\partial}

	\newcommand{\ft}{\tilde{f}}

	\newcommand{\cO}{\mathcal{O}}
	\newcommand{\cR}{\mathcal{R}}
	
	\newcommand{\be}{\begin{equation}}
		\newcommand{\ee}{\end{equation}}

\begin{document}

\begin{frontmatter}

\title{Evidence for a novel shift-symmetric universality class \\ from the functional renormalization group}

\author{Cristobal Laporte$^1$}
\ead{C.L.Munoz@science.ru.nl}
\author{Nora Locht$^1$}
\ead{C.Locht@student.science.ru.nl}
\author{Antonio D. Pereira$^{1,2}$}
\ead{adpjunior@id.uff.br}
\author{Frank Saueressig$^1$}
\ead{f.saueressig@science.ru.nl}

\address{$^1$ Institute for Mathematics, Astrophysics and Particle Physics (IMAPP) \\ Radboud University, Heyendaalseweg 135, 6525 AJ Nijmegen,The Netherlands \\[2ex]
	%
$^2$ Instituto de F\'isica, Universidade Federal Fluminese, \\
Av. Litor\^anea s/n, 24210-346, Niter\'oi, RJ, Brazil}

\begin{abstract}
Wetterich's equation provides a powerful tool for investigating the existence and universal properties of renormalization group fixed points exhibiting quantum scale invariance. Motivated by recent works on  asymptotically safe scalar-tensor theories, we develop a novel approximation scheme which projects the functional renormalization group equation onto functions of the kinetic term. Applying this projection to scalars and gauge fields, our analysis identifies a new universality class with a very special spectrum of stability coefficients. The implications of our findings in the context of asymptotically safe gravity-matter systems are discussed. 
\end{abstract}

\begin{keyword}
functional renormalization group, scalar field theory, gauge theory, universality, conformal field theory 
\end{keyword}

\end{frontmatter}

\section{Introduction}
\label{sect.intro}
Functional renormalization group (RG) techniques provide powerful tools for studying critical phenomena and potential high-energy completions of quantum field theories \cite{Berges:2000ew,Dupuis:2020fhh}. In particular, the Wetterich equation \cite{Wetterich:1992yh,Morris:1993qb,Reuter:1993kw,Reuter:1996cp}, capturing the scale-dependence of the effective average action $\Gamma_k$, constitutes a potent tool for studying the existence  of RG fixed points and their universal properties. Thus, it is not surprising that they have been applied to a vast range of physical systems covering statistical physics \cite{Dupuis:2020fhh}, particle physics \cite{Gies:2006wv,Pawlowski:2005xe}, and gravity \cite{Codello:2008vh,Reuter:2012xf,Eichhorn:2018yfc,Pawlowski:2020qer}. In particular, $O(N_\varphi)$-universality classes have been scrutinized in detail \cite{Balog:2019rrg,DePolsi:2020pjk}, yielding critical exponents in agreement with conformal bootstrap techniques \cite{Kos:2014bka,Chester:2019ifh}. In the context of quantum gravity the Wetterich equation has been the primary tool for investigating potential high-energy completions of gravity and gravity-matter systems within the gravitational asymptotic safety program \cite{Percacci:2017fkn,Reuter:2019byg}. 

The central idea of  the gravitational asymptotic safety program is that the short distance properties of gravity are controlled by a RG fixed point with non-vanishing interactions \cite{Weinberg:1980gg}: classical and quantum corrections balance in such a way that one obtains an enhanced symmetry, so-called quantum scale symmetry \cite{Wetterich:2019qzx,Wetterich:2020cxq}. As a consequence, physical quantities like scattering amplitudes remain finite at high energy \cite{Weinberg:1980gg}, also see \cite{Knorr:2019atm,Draper:2020bop,Platania:2020knd,Bonanno:2021squ,Fehre:2021eob,Knorr:2022lzn,Platania:2022gtt} for recent implementations and further discussions. At this stage there is substantial evidence that gravity indeed possesses a suitable RG fixed point -- the Reuter fixed point -- which gives rise to a consistent, predictive, and phenomenologically viable theory of quantum gravity, see  \cite{Gies:2016con,Denz:2016qks,Falls:2018ylp,Falls:2020qhj,Bonanno:2020bil,Knorr:2021slg,Baldazzi:2021orb,Sen:2021ffc} for further reading and additional references.

An intriguing consequence of the interacting nature of the gravitational fixed point arises upon including matter degrees of freedom. Since gravity couples to any type of matter, quantum fluctuations in the gravitational sector induce non-minimal couplings between matter fields and spacetime curvature \cite{Daas:2020dyo,Daas:2021abx,Eichhorn:2017sok} as well as matter self-interactions \cite{Eichhorn:2012va,Eichhorn:2016esv}. Structurally, these interactions are essential for guaranteeing a healthy behavior in gravity-mediated scattering processes \cite{Draper:2020bop}. The characteristic property of these interactions is that they respect the symmetries of the kinetic term. At the level of scalar-tensor theories, this has been cast into a non-renormalization theorem, stating that the space of action functionals compatible with these symmetries is closed with respect to the RG flow \cite{Laporte:2021kyp}. This inspires the investigation of RG flows in approximations projecting the exact functional renormalization group equation onto this subspace.

A quite peculiar property emerging within this context arose from studying the fixed point structure of four-dimensional scalar-tensor theories coupling gravity to scalar fields taking into account non-minimal interactions \cite{deBrito:2021pyi,Steinwachs:2021jft,Laporte:2021kyp,Knorr:2022ilz}. Decoupling the gravitational interactions suggested that there could actually be an interacting RG fixed point -- a so-called non-Gaussian fixed point (NGFP) -- complementing the free field theory. While the shift-symmetric projection of the RG flow does not exclude such a result per se, there are several reasons calling for a critical assessment of this situation: firstly, it is at variance with the general expectation that scalar field theory in $d=4$ dimensions admits only one renormalization group fixed point, the Gaussian fixed point (GFP). While for $d < 4$ one has the Wilson-Fisher fixed point \cite{Wilson:1971dc}, this fixed point merges with the Gaussian one at the upper critical dimension of this universality class $d=4$. Secondly, RG studies based on the local potential approximation (LPA) \cite{Wetterich:1992yh,Morris:1994ie,Berges:2000ew,Canet:2002gs,Litim:2002cf,Delamotte:2007pf}
do not yield evidence supporting the existence of this potentially new universality class. By construction, the LPA probes fixed points with different symmetry requirements though so that it is difficult to draw conclusions about the fixed point structure seen in the shift-symmetric projection on this basis. Finally, investigations of the RG flow of scalar-tensor theories employing a LPA-type scheme identified a GFP for the matter couplings only  \cite{Narain:2009fy,Percacci:2015wwa,Oda:2015sma,Ohta:2021bkc}. 

The goal of our letter is to clarify this puzzling situation. For this purpose, we study solutions of the Wetterich equation projected onto functions $f_k$ of the kinetic term $X$ constructed from a real scalar field, its extension to $O(N_\varphi)$-models, and Abelian gauge fields. We discover that all these systems converge against the same universality class once the polynomial approximation of $f_k(X)$ is pushed to a sufficiently high order. While the position of the underlying interacting fixed point approaches the free fixed point, the critical exponents of the new solution exhibit an intricate spectral pattern: the largest irrelevant coefficient flips its sign while all other coefficients converge to the classical mass dimension of the couplings contained in the approximation. To the best of our knowledge, this is the first time that this pattern of stability coefficients is discussed in detail.

\section{Wetterich equation and local potential approximation}
\label{sect.wetterich}
We start by reviewing the main concepts of the functional renormalization group and its fixed points in Sect.\ \ref{sect2.1} before illustrating their application in the context of scalar field theory using the local potential approximation in Sect.\ \ref{sect.2.2}.
\subsection{The functional renormalization group}
\label{sect2.1}
The Wetterich equation \cite{Wetterich:1992yh,Morris:1993qb,Reuter:1993kw,Reuter:1996cp} realizes the idea of the Wilsonian renormalization group \cite{Wilson:1973jj} at the level of the effective average action $\Gamma_k$. Schematically, the equation takes the form
\be\label{eq.frge}
	\p_t \Gamma_k = \frac{1}{2} {\rm Tr} \left[ \left( \Gamma_k^{(2)} + \cR_k \right)^{-1} \, \p_t \cR_k \right] \, . 
\ee
Here $t \equiv \ln(k/k_0)$ is the RG time, $\Gamma^{(2)}_k$ denotes the second functional derivative of $\Gamma_k$ with respect to the fluctuation fields and the trace contains an integration over loop momenta as well as a sum over fields. The infrared (IR) regulator $\cR_k$ provides a $k$-dependent mass term for fluctuations with momenta $p^2 \lesssim k^2$ and vanishes for $p^2 \gg k^2$. The interplay of the regulators appearing on the right-hand side of \eqref{eq.frge} renders the trace finite in the UV and IR. Its main contribution stems from fluctuations with momenta $p^2 \simeq k^2$, indicating that the flow of $\Gamma_k$ is actually driven by integrating out fluctuations with momenta comparable to the coarse graining scale $k$. In explicit computation we will chose a regulator of Litim-type \cite{Litim:2000ci,Litim:2001up}
\be\label{eq.litimreg}
R_k(p^2) = (k^2 - p^2) \, \Theta(1-p^2/k^2) \, 
\ee
with $\Theta(x)$ the Heaviside step function.

The Wetterich equation lives on theory space, the space containing all action functionals $\cO$ which can be constructed from a given field content and compatible with symmetry requirements. Non-perturbative computations based on this framework typically project the exact RG flow on a subspace of theory space. In practice, this is implemented by truncating the action functionals in $\Gamma_k$ typically to a finite set
\be\label{eq.trunc}
\Gamma_k \simeq \sum_{i=1}^N \, \bar{u}_i(k) \, \cO_i \, . 
\ee
Substituting this ansatz into \eqref{eq.frge} and comparing the coefficients multiplying the monomials $\cO_i$ allows to read off the beta functions encoding the running of the couplings $\bar{u}_i(k)$. Converting to dimensionless couplings $u_i = \bar{u}_i \, k^{-d_i}$ where $d_i$ is the mass-dimension of $\bar{u}_i$, these beta functions take the form of a coupled system of autonomous, first-order differential equations
\be\label{eq.betadef}
\p_t u_i = \beta_{u_i}(\{u_j\}) \, , \qquad i \in 1,\cdots, N \, . 
\ee
Fixed points $\{u_i^*\}$ appear as roots of this system of equations $\beta_{u_i}(\{u_j^*\}) = 0$, $i=1,\cdots,N$. In the vicinity of a fixed point the properties of the RG flow can be studied by linearizing the beta functions
\be\label{eq.linflow}
\p_t u_i \approx \sum_j {\bf B}_i{}^j \, (u_j - u_j^*) \, . 
\ee
Here ${\bf B}_i{}^j \equiv \frac{\p}{\p u_j} \beta_{u_i}|_{u_i = u_i^*}$ is the stability matrix at the fixed point. Its critical exponents $\theta_i$, defined as minus the eigenvalues of ${\bf B}_i{}^j$, encode whether the corresponding eigendirection is UV-attractive (Re($\theta> 0$)), UV-repulsive (Re($\theta  < 0$)), or marginal (Re($\theta$) = 0). For a Gaussian fixed point, corresponding to a free theory, the critical exponents are given by the canonical mass dimension of the coupling.

A healthy UV-completion arises when a RG trajectory approaches a fixed point as $k \rightarrow \infty$. In other words, RG trajectories whose high-energy completion is provided by a given fixed point must emanate from this fixed point along a UV-attractive direction. The number of these directions equals the number of free parameters in the construction. Consequently, the predictive power of a fixed points increases with a lower number of UV-relevant directions.  
\subsection{RG fixed points for scalar field theory}
\label{sect.2.2}
We now proceed by studying the RG fixed points for a four-dimensional scalar field theory, based on the local potential approximation. 
We start from a single, real scalar field $\phi$ in a $d$-dimensional Euclidean spacetime. The local potential approximation retains a scale-dependent scalar potential, approximating
\be\label{eq.lpa}
\Gamma_k[\phi] \simeq \int d^dx \left( \frac{1}{2} (\p_\mu \phi) (\p^\mu \phi) + V_k[\phi] \right) \, . 
\ee
In principle, it is straightforward to also include the wave-function renormalization for $\phi$ \cite{Tetradis:1993ts,Delamotte:2007pf}. At this stage this is not needed though. Following \cite{Reuter:2019byg}, evaluating \eqref{eq.frge} in the LPA gives a partial differential equation capturing the scale-dependence of $V_k[\phi]$. This equation is readily obtained by taking the second functional derivative of the ansatz and evaluating the right-hand side of \eqref{eq.frge} for the case where $\phi$ is constant. The result reads
\be\label{eq.genfct1}
\p_t V_k[\phi] = \frac{1}{2} \int \frac{d^dp}{(2\pi)^d} \, \left( p^2 + R_k + V_k^{\prime\prime}[\phi] \right)^{-1} \,\p_t R_k \, . 
\ee
Here the prime denotes a derivative of $V_k[\phi]$ with respect to the scalar field. Since the integrand  is a function of $p^2$ only, carrying out the angular integrations is straightforward,
\be\label{eq.genfct2}
\p_t V_k[\phi] = \frac{v_d}{2} \int_0^\infty d\omega \, \omega^{d/2-1} \, \left( \omega + R_k + V_k^{\prime\prime}[\phi] \right)^{-1} \,\p_t R_k \, , 
\ee
where $v_d = ((4\pi)^{d/2} \Gamma[d/2])^{-1}$ is the angular volume factor. In order to perform the final integral, we specify the regulator function to \eqref{eq.litimreg}, yielding
\be\label{eq.genfct3}
\p_t V_k[\phi] = \frac{2 v_d}{d}  \, \frac{k^{d+2}}{k^2 + V_k^{\prime\prime}[\phi] } \, . 
\ee

This result serves as a generating functional for the beta functions appearing in a polynomial expansion of $V_k[\phi]$,
\be
V_k[\phi] = \sum_{n=1}^N \, \bar{u}_n(k) \, (\phi^2)^n \, . 
\ee 
Here $\bar{u}_1 = \frac{1}{2} m^2$ encodes the mass of the field and $\bar{u}_2$ corresponds to the $\phi^4$ interaction. Substituting this expansion into \eqref{eq.genfct3} and equating equal powers of $\phi^2$ leads to the beta functions for the dimensionful couplings. Converting to the dimensionless couplings $u_n \equiv (k^2)^{n-2} \bar{u}_n$ and setting $d=4$, the set of equations obtained at order $N=2$ reads
\be\label{eq.betalpa}
\begin{split}
\p_t u_1 = & \, - 2 \, u_1 - \frac{3 u_2}{8 \pi^2 (1 + 2 u_1)^2 } \, , \\
\p_t u_2 = & \, 
 \frac{9 u_2^2}{2 \pi^2 (1+2 u_1)^3} \, . 
\end{split}
\ee
Higher-order truncations are readily generated using computer algebra software. Obviously, the only fixed point of \eqref{eq.betalpa} is the GFP $u_1^* = 0, u_2^* = 0$. Extending the analysis to include further terms shows that this is the only stable fixed point solution. In this case the critical exponents are given by $\theta_n^{\rm GFP} = 2n- 4$.
\section{Flow equations in the shift-symmetric approximation}
\label{sect.ssa}
In the shift-symmetric approximation (SSA) the effective average action is projected to a function of the kinetic term $X$, i.e.,
\be\label{eq.ssa}
\Gamma_k[\phi] \simeq \int d^dx \, f_k(X) \, , 
\ee
We derive the flow equation satisfied by $f_k(X)$ in the context of scalar fields (Sect.\ \ref{sect.ssascalar}), $O(N_\varphi)$-models (Sect.\ \ref{sect.onmodels}), and Abelian gauge fields (Sect.\ \ref{sect.photons}). Our main results are the generating functionals \eqref{eq.dtfk4}, \eqref{o(n).9}, and \eqref{mainres.gauge} which encode the scale-dependence of $f_k(X)$ in these cases.
\subsection{Scalar field theory}
\label{sect.ssascalar}
For a real scalar field $\phi$, the kinetic term is given by
\be\label{eq.Xdef}
X = \frac{1}{2} \, Z_k (\p_\mu \phi) (\p^\mu \phi) \, . 
\ee
Here $Z_k$ is the wave function renormalization and we define the scalar anomalous dimension
\be\label{eq.etadef}
\eta_s = - \p_t \ln Z_k \, . 
\ee
At this stage, it is convenient to decompose $f_k(X)$ into the kinetic term and interactions $\tilde{f}(X)$,
\be\label{eq.fkans}
f_k(X) = X + \tilde{f}_k(X) \, .
\ee
The Hessian of $\Gamma_k$ resulting from \eqref{eq.ssa} is
\be\label{eq.hessianssa}
\Gamma_k^{(2)}(p) = Z_k \left[ p^2 + \ft^\prime_k \, p^2 + Z_k \, \ft^{\prime\prime}_k \, p^\mu p^\nu (\p_\mu \phi) (\p_\nu \phi) \right] \, . 
\ee
Here the primes denote derivatives with respect to $X$ and we set $(\p_\mu \phi)$ to a constant. Similarly to the LPA, this suffices to project the RG flow onto the subspace spanned by \eqref{eq.fkans}. The evaluation of the Wetterich equation for this ansatz then proceeds as follows. Including the wave function renormalization in the regulator by setting $\cR_k = Z_k R_k$ and substituting \eqref{eq.hessianssa} into \eqref{eq.frge} yields
\be\label{eq.dtfk1}
\begin{split}
\p_t f_k = \frac{1}{2 Z_k} &  \int \frac{d^dp}{(2\pi)^d}  \,  \p_t \cR_k \\ \times & \left[ P_k + \ft^\prime p^2 + \ft^{\prime\prime} Z_k p^\mu p^\nu (\p_\mu \phi) (\p_\nu \phi) \right]^{-1}   \, , 
\end{split}
\ee
where $P_k \equiv p^2 + R_k$. In order to simplify the structure of the momentum integral, we expand the Hessian in a geometric series
\be\label{eq.dtfk2}
\begin{split}
	\p_t f_k = & \frac{1}{2} \int \frac{d^dp}{(2\pi)^d} \, Z_k^{-1} \p_t \cR_k \\ & \times \left(\sum_{n=0}^\infty \left[ P_k + \ft^\prime p^2 \right]^{-(n+1)} \left[ - \ft^{\prime\prime} Z_k p^\mu p^\nu (\p_\mu \phi) (\p_\nu \phi) \right]^{n} \right) \, . 
\end{split}
\ee
Since $\p_\mu\phi$ is a book-keeping device which can, in principle, be chosen to be infinitesimal, one can guarantee the convergence of the series. Using the identity \eqref{eq.ssamaster} allows to replace the products of uncontracted momenta $p^\mu p^\nu$ by the square of the momentum
\be\label{eq.dtfk3}
\begin{split}
	\p_t f_k = & \frac{1}{ Z_k}\int \frac{d^dp}{(2\pi)^d} \, \p_t \cR_k \, \left[ P_k + \ft^\prime p^2 \right]^{-1} \\ & 
	\times \left(\sum_{n=0}^\infty \frac{\Gamma(d/2) }{\Gamma(d/2+n)} \, \frac{\Gamma(2n)}{ \Gamma(n)} \left( - \frac{X \ft^{\prime\prime} p^2}{2(P_k + \ft^\prime p^2)} \right)^n \right)\, . 
\end{split}
\ee
Notably, the series can be resumed. Keeping $d$ general, this results in a hypergeometric function. Restricting to $d=4$, this result simplifies significantly, with {\tt Mathematica} yielding
\be\label{eq.dtfk4}
\begin{split}
	\p_t f_k = & \frac{1}{2 Z_k} \int \frac{d^4p}{(2\pi)^4} \, \frac{ \p_t \cR_k}{X \ft^{\prime\prime} p^2} \, \left(\sqrt{1+ \frac{2 X \ft^{\prime\prime} p^2}{P_k + \ft^\prime p^2}} - 1 \right). 
\end{split}
\ee
This result is the analogue of eq.\ \eqref{eq.genfct1} for the SSA. Note that, despite the occurrence of a factor $1/X$, the argument of the momentum integral is actually regular at $X=0$ once the square root is expanded for small values $X$.

In order to progress further, we specify the regulator $\cR_k = Z_k R_k$ with the scalar profile given in eq.\ \eqref{eq.litimreg}. This implies
\be
Z_k^{-1} \p_t \cR_k = (2 k^2 - \eta_s (k^2-p^2)) \Theta(k^2-p^2) \, , \quad P_k = k^2\, . 
\ee
This choice allows to carry out the momentum integral in \eqref{eq.dtfk4} analytically. As a result we obtain the projected flow equation governing the scale-dependence of $f_k(X)$:
\be\label{eq.masterssa}
\begin{split}
8 \pi^2 & \, k^{-4}   \p_t f_k =  - \frac{1+\ft^\prime}{2 X \ft^\prime \ft^{\prime\prime}} (1-b) \\
& + \frac{\ft^\prime \, (1+ \ft^\prime)^2 +  X \ft^{\prime\prime} \left(3  + 4 \ft^{\prime} + 2 (\ft^\prime)^2 \right)}{8 X (\ft^\prime)^2 \ft^{\prime\prime} (\ft^\prime + 2 X \ft^{\prime\prime}) } \eta_s \\  
& - \frac{(1+\ft^\prime) (\ft^\prime(1+\ft^\prime) + X \ft^{\prime\prime} (3+2\ft^\prime)  )}{8 X (\ft^\prime)^2 \ft^{\prime\prime} (\ft^\prime + 2 X \ft^{\prime\prime})} \, b \, \eta_s \\
& +  \frac{1}{ (\ft^\prime)^2} \, \left(  1 - \frac{2 \ft^\prime (1+\ft^\prime) +  X \ft^{\prime\prime} (3 + 4 \ft^\prime)}{4 \ft^\prime \, (\ft^\prime + 2 X \ft^{\prime\prime})} \eta_s \right) \\ & \qquad \times a \, \left({\rm arctanh}(ab) -  {\rm arctanh}(a) \right) \, . 
\end{split}
\ee
Here we abbreviated
\be
\begin{split}
a & \equiv \left( \frac{\ft^\prime}{\ft^\prime + 2 X \ft^{\prime\prime}} \right)^{1/2} \, , \quad 
b  \equiv  \left( \frac{1+ \ft^\prime + 2 X \ft^{\prime\prime}}{1+\ft^\prime } \right)^{1/2} \, . 
\end{split}
\ee
This constitutes the main result of the SSA in the case of scalar field theory.

\subsection{$O(N_{\varphi})$-models}
\label{sect.onmodels}
The natural extension of the previous section considers the case of $N_{\varphi}$ scalar fields. In this case, the kinetic term is a straightforward generalization of \eqref{eq.Xdef},
\begin{equation}
\mathbb{X} = \frac{Z_k}{2}(\p_\mu\varphi^a)(\p^\mu\varphi^a)\,,
\label{o(n).1}
\end{equation}
with $a=1,\ldots , N_{\varphi}$. Demanding $O(N_\varphi)$-invariance next to shift symmetry give rise to a rich set of allowed interactions. Defining the tensor
\begin{equation}
\mathbb{X}^{ab}_{\mu\nu} = \frac{Z_k}{2}(\p_\mu\varphi^a)(\p_\nu\varphi^b)\,,
\label{o(n).2}
\end{equation}
and its partial trace $\mathbb{X}_{\mu\nu} \equiv \delta^{ab}\mathbb{X}^{ab}_{\mu\nu}$, it is straightforward to verify that these organize themselves in three classes:
\begin{enumerate}
\item Powers of single-full traces of $\mathbb{X}^{ab}_{\mu\nu}$, i.e., $\mathbb{X}^n$, for some natural $n$;
\item Single-traces of strings of $\mathbb{X}_{\nu_1}{}^{\nu_2}\mathbb{X}_{\nu_2}{}^{\nu_3}\ldots \mathbb{X}_{\nu_n}{}^{\nu_1}$, $n \ge 2$, and powers of them;
\item Products of interactions of classes 1 and 2. 
\end{enumerate}
In terms of truncations of the effective average action, the analogue of \eqref{eq.fkans} comprising all the $O(N_\varphi)$- and shift-symmetric interactions is
\begin{equation}
\Gamma_k [\varphi^a] \simeq \int d^dx~F_k (\mathbb{X},\mathbb{X}_{\mu\nu})\,.
\label{o(n).3}
\end{equation}
Introducing $F_k (\mathbb{X},\mathbb{X}_{\mu\nu}) = \mathbb{X} + \tilde{F}_k (\mathbb{X},\mathbb{X}_{\mu\nu})$ leads to
\be
\begin{split}
\Big({\Gamma}^{(2)}_k\Big)^{ab} = & \, Z_k \delta^{ab}p^2 \\
&+Z_k \delta^{ab}\Big[\tilde{F}^{(1,0)}_k\,p^2+\Big(\tilde{F}^{(0,1)}_k\Big)_{\mu\nu}p^\mu p^\nu\Big]
\\
&+ 2Z_k \Big[\tilde{F}^{(2,0)}_k \mathbb{X}^{ab}_{\mu\nu} 
+ \Big(\tilde{F}^{(0,2)}_k\Big)_{\mu\alpha,\nu\beta}\mathbb{X}^{ab,\alpha\beta}\Big]p^\mu p^\nu
\\
&+ 4Z_k \Big(\tilde{F}^{(1,1)}_k\Big)_{\mu\nu} \mathbb{X}^{ab,\nu\alpha}p^\mu p_\alpha\,,
\label{o(n).4}
\end{split}
\ee
with
\begin{equation}
\tilde{F}^{(n,m)}_{\mu_1 \nu_1 , \ldots ,\mu_m \nu_m} \equiv \frac{\partial^{n+m}\tilde{F}_k}{\partial \mathbb{X}^n \partial {\mathbb{X}}_{\mu_1 \nu_1}\ldots \partial {\mathbb{X}}_{\mu_m \nu_m}}\,.
\label{o(n).5}
\end{equation}
For the purposes of this work, we choose a truncation that encompasses just interactions of type 1, i.e., powers of single-full traces. Hence, $\tilde{F}_k (\mathbb{X},\mathbb{X}_{\mu\nu}) \simeq \tilde{f}_ k (\mathbb{X})$. It is important to emphasize that the truncation $F_{k} (\mathbb{X},\mathbb{X}_{\mu\nu}) = f_k (\mathbb{X})$ is not self-consistent in terms of its tensorial structure. Hence, the series expansion in \eqref{o(n).8} takes into account the projection to the appropriate basis of the chosen truncation. Moreover, the limit $N_\varphi \to 1$ does not recover the single-scalar result. This is expected since one has discarded tensorial structures which are indistinguishable in the single-scalar model.

The Hessian \eqref{o(n).4} then simplifies to
\begin{eqnarray}
	\Big({\Gamma}^{(2)}_k\Big) &=& Z_k \Big[ p^2(1+\tilde{f}^{\prime}_k) \mathbf{1} + 2\tilde{f}^{\prime\prime}_k\,p \cdot [\mathbb{X}]\cdot p \Big]\,.
	\label{o(n).6}
\end{eqnarray}
where $\mathbf{1}$ is the unit on field space, satisfying tr($\mathbf{1}) = N_\varphi$, and $p \cdot [\mathbb{X}]\cdot p \equiv p^\mu \, [\mathbb{X}]_{\mu\nu}^{ab} \, p^\nu$. At this point we have all the elements to extract the flow of $f_k (\mathbb{X})$. Specifying the regulator $\mathbb{R}_k \equiv \cR_k \, \mathbf{1}$, the flow of $f_k$ is
\be
\begin{split}
\partial_t f_k = \frac{1}{2Z_k}&{\rm tr} \int \frac{d^dp}{(2\pi)^{d}}\partial_t \mathbb{R}_k
\\ \times & \Big[\mathbf{1}\, p^2(1+\tilde{f}^{\prime}_k)+ 2\tilde{f}^{\prime\prime}_k\, p \cdot [\mathbb{X}]\cdot p\Big]^{-1}\,,
\label{o(n).7}
\end{split}
\ee
with ${\rm tr}$ denoting the trace over the internal indices. This equation can be expanded as
\be
\begin{split}
\partial_t f_k &= \frac{1}{2Z_k}\int \frac{d^dp}{(2\pi)^{d}}\partial_t \cR_k \frac{N_\varphi}{P_k + \tilde{f}^\prime p^2}
\\ &+ \frac{1}{2Z_k} \int \frac{d^dp}{(2\pi)^{d}}\partial_t \cR_k \sum^{\infty}_{n=1}\frac{(-\tilde{f}^{\prime\prime}_k p^2 \mathbb{X})^n}{(P_k + \tilde{f}_k p^2)^{n+1}}\frac{\Gamma(d/2)}{\Gamma(d/2+n)}\,.
\label{o(n).8}
\end{split}
\ee
  As in the single-scalar model, the series above can be resumed. In $d=4$,
\be
\begin{split}
\partial_t f_k = &\frac{1}{2Z_k}\int \frac{d^4p}{(2\pi)^{4}}\partial_t \cR_k 
\frac{\frac{f^{\prime\prime}_k (N_\varphi -1) p^2 \mathbb{X}}{P_k+f^{\prime}_k p^2}-{\rm exp}\Big(-\frac{f^{\prime\prime}_k p^2 \mathbb{X}}{P_k+f^\prime_k p^2}\Big)+1}{ f^{\prime\prime}_k p^2 \mathbb{X}}\,.
\label{o(n).9}
\end{split}
\ee
Specifying the regulator as in the previous subsection, the integral in \eqref{o(n).9} can be performed analytically, but the resulting expression is not particularly illuminating and will not be reported here.
\subsection{Abelian gauge fields}
\label{sect.photons}
The SSA for an Abelian gauge field $A_\mu$ can be developed analogously to the scalar case. We define the field strength tensor $F_{\mu\nu} \equiv \p_\mu A_\nu - \p_\nu A_\mu$ and  denote the kinetic term for the gauge field by
\be
X \equiv \frac{Z_k}{4} F_{\mu\nu} F^{\mu\nu} \, . 
\ee
Here $Z_k$ is the wave-function renormalization of the gauge field and we introduce its anomalous dimension $\eta_a \equiv - \p_t \ln Z_k$. The SSA-ansatz for the effective average action then consists of the kinetic term, a gauge-fixing term for which we adopt Feynman gauge and interactions $\tilde{f}(X)$:
\be\label{ans.gauge}
\Gamma_k[A_\mu] \simeq \int d^dx \left\{ X + \frac{Z_k}{2} (\p_\mu A^\mu)^2 + \tilde{f}(X) \right\} \, . 
\ee
Since the ghost sector is independent of the gauge field it does not contribute to the present computation and can be ignored. We then project the flow onto the ansatz spanned by \eqref{ans.gauge}.

The Hessian resulting from \eqref{ans.gauge} now carries Lorentz indices and reads
\be\label{hessian.gauge}
\left[\Gamma_k^{(2)}\right]_\mu{}^\nu = Z_k \left[ p^2 (1 + \tilde{f}_k^\prime ) \delta_\mu^\nu -  \tilde{f}_k^\prime p_\mu p^\nu + Z_k \tilde{f}_k^{\prime\prime} F_\mu{}^\alpha F^{\nu\beta} p_\alpha p_\beta \right] \, . 
\ee
 In order to track the $X$-dependence, it suffices to work with a constant field strength tensor $F_{\mu\nu}$. Including the wave-function renormalization in the regulator by setting 
 $\left[\mathbb{R}_k\right]_{\mu}{}^\nu \equiv \cR_k \, \delta_{\mu}{}^\nu$, we arrive at
\be\label{eq.dtfk4gf}
\begin{split}
	\p_t f_k = \frac{1}{2 Z_k} &  {\rm tr} \int \frac{d^dp}{(2\pi)^d}  \p_t \mathbb{R}_k \\ \times & \left[ (P_k + p^2 \tilde{f}_k^\prime) \delta_\mu^\nu -  \tilde{f}_k^\prime p_\mu p^\nu + Z_k \tilde{f}_k^{\prime\prime} F_\mu{}^\alpha F^{\nu\beta} p_\alpha p_\beta  \right]^{-1}
\end{split}
\ee
where tr is a sum over internal indices. In order to progress further, it is again useful to expand the inverse on the right-hand side in terms of a geometric series,
\be\label{eq.dtfk4geo}
\begin{split}
	\p_t f_k = \frac{1}{2 Z_k} &  {\rm tr} \int \frac{d^dp}{(2\pi)^d}  \p_t \mathbb{R}_k 
	\\ \times &
	\sum_{n=0}^\infty  (-1)^n \, \left[ P_k + p^2 \tilde{f}_k^\prime \right]^{-(n+1)}\left[ A_\mu{}^\nu + B_\mu{}^\nu \right]^n \, , 
\end{split}
\ee
where
\be
\begin{split}
	A_\mu{}^\nu \equiv & \, - \tilde{f}_k^\prime p_\mu p^\nu \, , \qquad
	B_\mu{}^\nu \equiv  Z_k \tilde{f}_k^{\prime\prime} F_\mu{}^\alpha F^{\nu\beta} p_\alpha p_\beta \, . 
\end{split}
\ee
These definitions make it explicit that the binomial appearing in \eqref{eq.dtfk4geo} actually reduces to two terms since owed to the antisymmetry of the field strength tensor any cross-product $A \cdot B = 0$. We then note that 
\be\label{series1}
{\rm tr}A^n = (- \tilde{f}_k^\prime \, p^2 )^n \, .
\ee
In order to evaluate the term proportional to tr$B^n$, we again make use of the substitution rule for loop-momenta \eqref{eq.prepgen}. This leads to the identity
\be\label{series2}
{\rm tr} \, B^n \, \mapsto  \, \frac{\Gamma(d/2)}{\Gamma(d/2+n)} \left( 2 X \,  \tilde{f}_k^{\prime\prime} \, p^2 \right)^n \, . 
\ee 
Note that in this expression we have retained terms containing the field strength in the form of $X$ only. Strings of contracted $F$'s with more than two powers of the field strength (like $F_\mu{}^\nu F_\nu{}^\rho F_\rho{}^\sigma F_\sigma{}^\mu)$ do not contribute to the SSA-truncation and have not been included in \eqref{series2}. 

We now restrict to $d=4$. Remarkably, both series arising from substituting eqs.\ \eqref{series1} and \eqref{series2} into \eqref{eq.dtfk4geo} can be resummed. This gives the generating function encoding the scale-dependence of $f_k$ in the case of an Abelian gauge field
\be\label{mainres.gauge}
\begin{split}
\p_t f_k = & \, \frac{1}{2 Z_k} \int \frac{d^4p}{(2\pi)^4} \, \p_t \cR_k 
 \left[ \frac{1}{P_k} + \frac{1-\exp\left( - \frac{2 X \tilde{f}^{\prime\prime} p^2 }{P_k + \tilde{f}^\prime p^2} \right)}{2 \, X \tilde{f}^{\prime\prime} p^2} \right] \, . 
\end{split}
\ee
The following observations are in order. Firstly, the first term in eq.\ \eqref{mainres.gauge} is independent of $X$. Hence it does not contribute to the flow of $f_k$. Secondly, we can again specify the regulator to be of the form \eqref{eq.litimreg}. With this choice the momentum-integral can be carried out analytically. The result is expressed in terms of exponential integrals $Ei(x)$. Again the result is rather lengthy and does not lend it self to a further analysis, we refrain from giving the corresponding formula here.

\section{Fixed points and universality classes}
\label{sect.universality}
We now determine the fixed points of the generating functionals \eqref{eq.dtfk4}, \eqref{o(n).9}, and \eqref{mainres.gauge}. Following the strategy employed by the LPA, we approximate $\ft(X)$ by a $N$th order polynomial starting at order $X^2$:
\be\label{polynomial}
\tilde{f}_k(X) \simeq \sum_{n=2}^N \bar{u}_n \, X^n \, . 
\ee
Converting to dimensionless couplings, we have
\be
\p_t f_k(X) = - \eta_i X + \sum_{n=2}^N \left(\beta_{u_n} + 4(1-n) u_n - n \eta_i u_n \right) k^{4(1-n)} X^n \, , 
\ee
where $i=s,\varphi,a$ for scalars, the $O(N_\varphi)$-model with $N_\varphi = 2$, and gauge fields, respectively. Substituting this expression into \eqref{eq.masterssa} and equating equal powers of $X$ on both sides gives a system of equations determining $\eta$ as well as $\beta_{u_n}$, $n=2,\cdots,N$.

Our main interest is in the fixed points of these beta functions and their stability properties. All systems admit a Gaussian fixed point, $u_n^* = 0, n \ge 2$, where the stability coefficients are given by the canonical mass-dimension of the coupling
\be
\theta_n^{\rm GFP} = -4n \, , \qquad n \ge 1 \, . 
\ee
We are then interested in the additional fixed points of the system.
 The anomalous dimension is obtained from evaluating \eqref{eq.dtfk4}, \eqref{o(n).9}, and \eqref{mainres.gauge} at linear order in $X$. This yields
\be\label{anomalousdim}
\begin{split}
\eta_s = & \frac{8 u_2}{ 128\pi^2 + u_2} \, , \\
\eta_\varphi = & \frac{8(1+2N_\varphi)u_2}{384 \pi^2 + (1+2N_\varphi)u_2} \, , \\
\eta_a = & \frac{8 u_2}{96 \pi^2 + u_2} \, . 
\end{split}
\ee
The structure of the beta functions arising from the coefficients $X^n$ for $n \ge 2$ turns out to be rather simple. In particular the fixed point condition $\beta_{u_n} = 0$ can be solved recursively employing completely analytic methods. At $N=2$ the system admits three fixed points. For the three systems, these are located at
\be\label{fp.seed}
\begin{array}{lccc}
	&   {\rm \quad scalar \quad}  & {\quad O(2) \quad} &  {\rm \quad gauge \; field \quad} \\
{\rm GFP:} & u_2^* = 0 & u_2^* = 0 & u_2^* = 0 \\
{\rm NGFP}_1: & u_2^* = -128 & u_2^* = -87 & u_2^* = -91.2 \\
{\rm NGFP}_2: & u_2^* = -12506 & u_2^* = -8690 & u_2^* = -8953 \\
\end{array}
\ee
The $O(N_\varphi)$-model displays two NGFPs for any $N_\varphi$. Curiously, the fixed-point value $u_2^*$ goes to zero as $N_\varphi \to \infty$, although the critical exponents are rather insensitive to $N_{\varphi}$. Thus besides the GFP, we obtain two candidates for NGFPs. It then turns out that NGFP$_2$ appears due to the inclusion of the anomalous dimension and vanishes in approximations where $\eta_i = 0$. This leaves NGFP$_1$ as a robust candidate. 

We follow this solution, increasing the dimension of the truncation subspace up to $N=15$.\footnote{Increasing $N$ introduces additional zeros in the system of beta functions. All these solutions can be tracked analytically. In contrast to the family of solutions originating from the seed NGFP$_1$, these do not exhibit obvious convergence patterns though. Therefore we do not discuss these roots in the sequel.} This analysis reveals the following striking property: the fixed points forming the family NGFP$_1$ actually behave almost identical in all three systems. This suggests that \emph{they belong to the same universality class}. This is highly remarkable since the fixed points appear in systems with different field content and number of degrees of freedom. This also entails that it is actually feasible to discuss their properties in a joined way. We make the following observations
\begin{figure}[h!]
	\includegraphics[width = 0.48\textwidth]{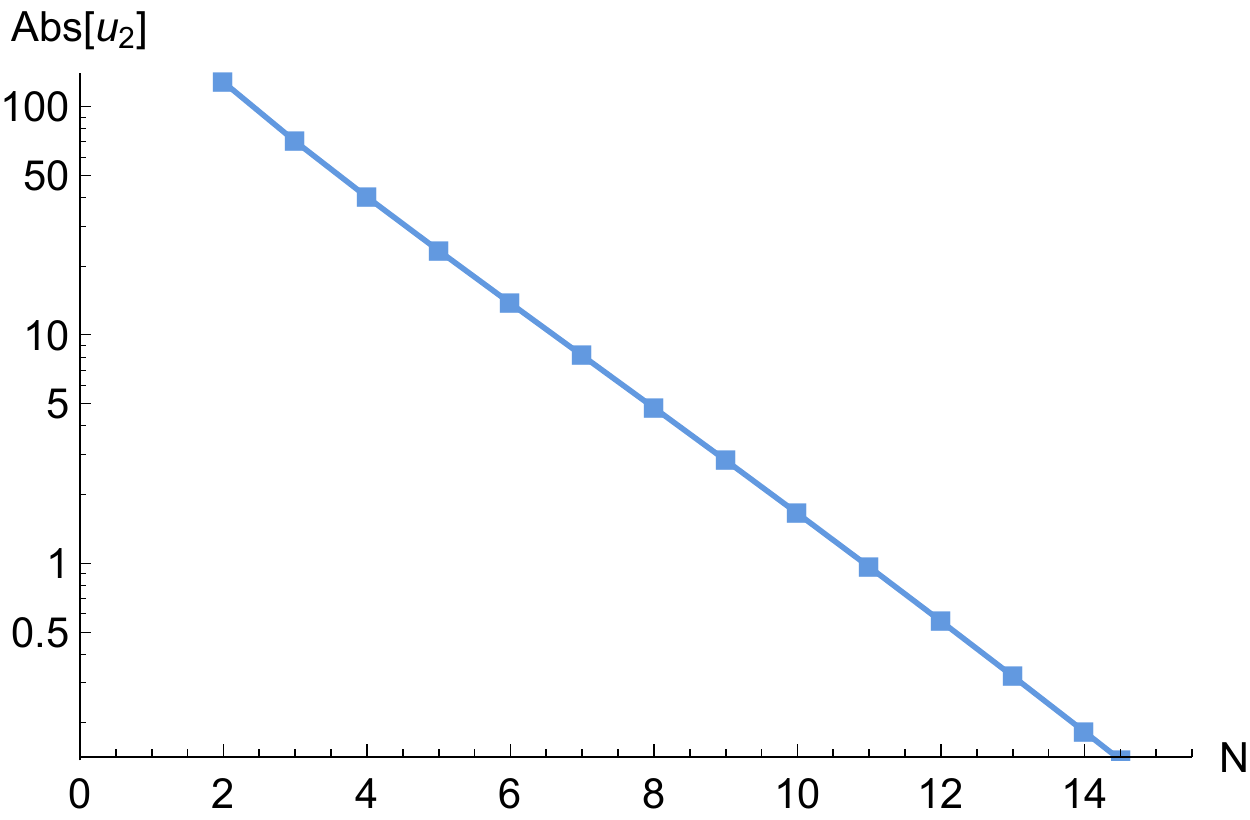} \\
	\includegraphics[width = 0.48\textwidth]{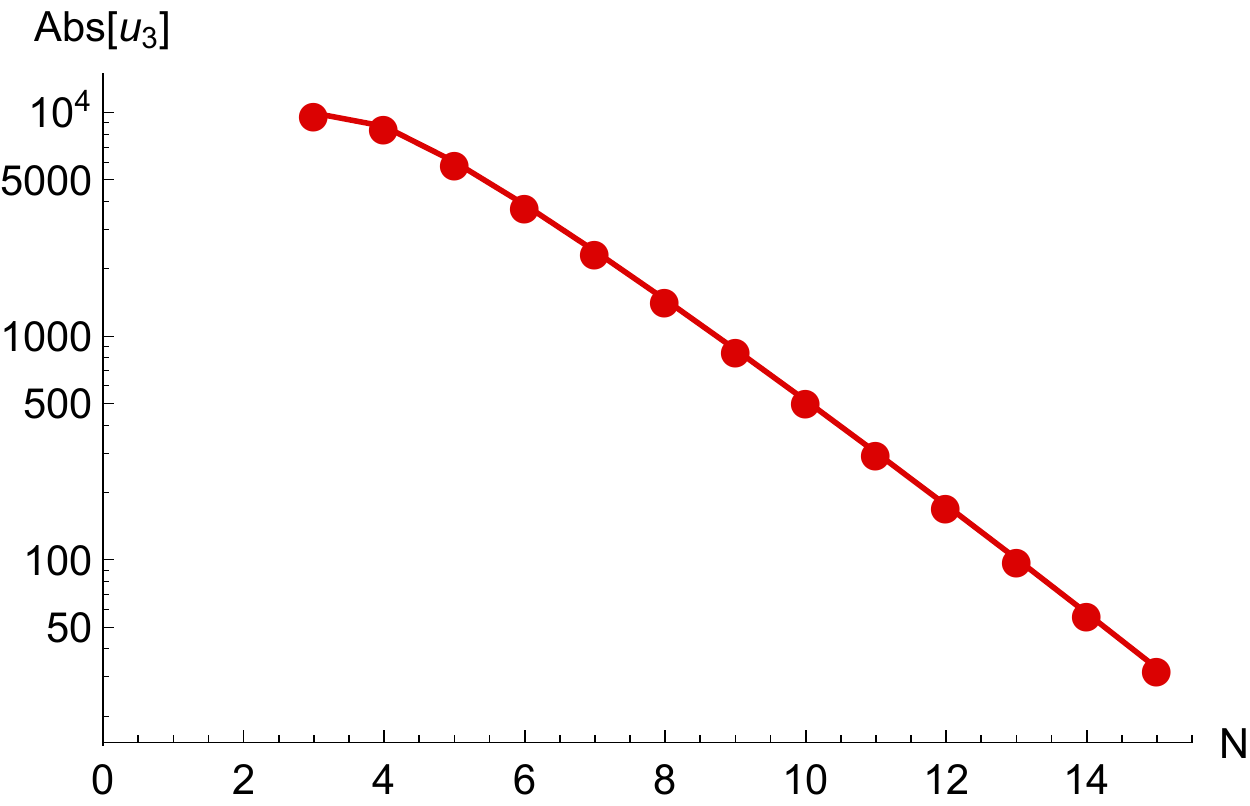} \\
	\includegraphics[width = 0.48\textwidth]{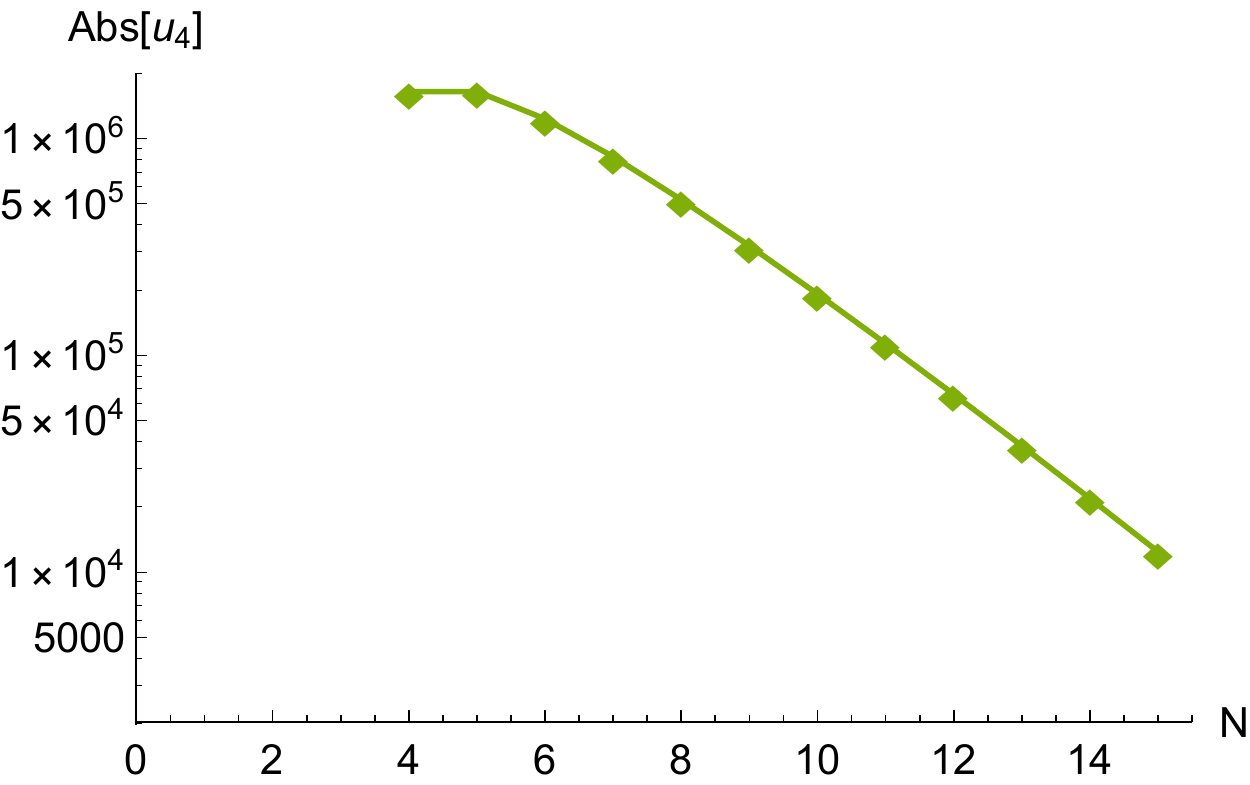} \\
	\caption{\label{fig.move} Position of the lowest order couplings $u_n^*$, $n=2,3,4$ for the NGFP$_1$ obtained from the scalar system as a function of $N$. The $O(N_\varphi)$-system and the gauge field exhibit a behavior identical to the one illustrated in the plots.}
\end{figure}
\begin{enumerate}
	\item Increasing $N$ the position of the NGFP$_1$ changes rapidly. In particular the values $u_n^*$ move towards zero. This shift is compensated by the new couplings appearing in the extended truncation taking successively larger fixed point values before moving towards the GFP. This behavior is exemplified based on the scalar system in Fig.\ \ref{fig.move}. 
	\item  As a consequence of $u_2^*$ approaching zero, the anomalous dimension $\eta_i$ associated with the fixed point goes to zero from below. This is illustrated in Fig.\ \ref{fig.etas}.
	\item While the position of the NGFP$_1$  changes by several orders of magnitude, its critical exponents converge rapidly. For the first three stability coefficients, this convergence is shown in Fig.\ \ref{fig.3}, which has been obtained in the approximation $\eta_i=0$. There is one UV-relevant eigendirection where $\theta_1 = 4$. All other stability coefficients come with a negative real part Re($\theta_n) < 0$, $n \ge 2$. For large $N$, $\theta_2$ and $\theta_3$ converge to the stability coefficients of the GFP.
\end{enumerate}

\begin{figure}[h!]
	\includegraphics[width = 0.48\textwidth]{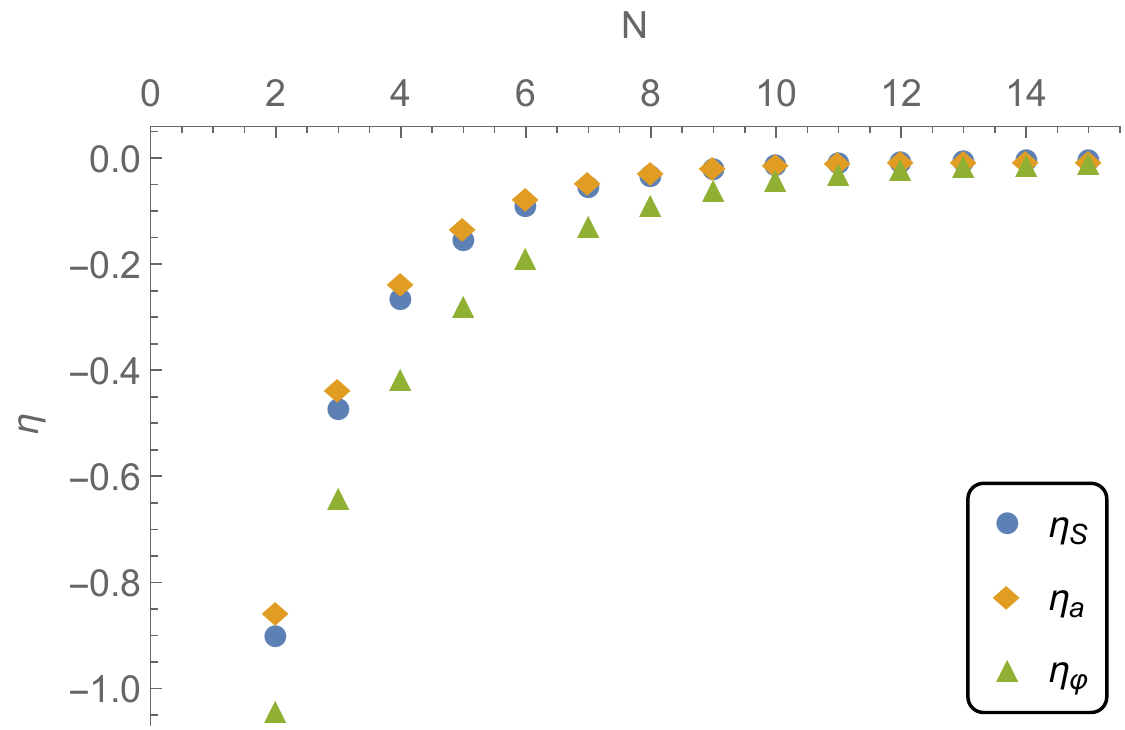} \\
	\caption{\label{fig.etas} $N$-dependence of the anomalous dimension $\eta_i$ at the NGFP$_1$ obtained for the scalar field (blue circles), Abelian gauge field (orange diamonds), and the $O(N_\varphi)$-model with $N_\varphi = 2$ (green triangles) for increasing truncation size $N$. The anomalous dimension falls of exponentially with increasing $N$.}
\end{figure}

\begin{figure}[h!]
	\includegraphics[width = 0.48\textwidth]{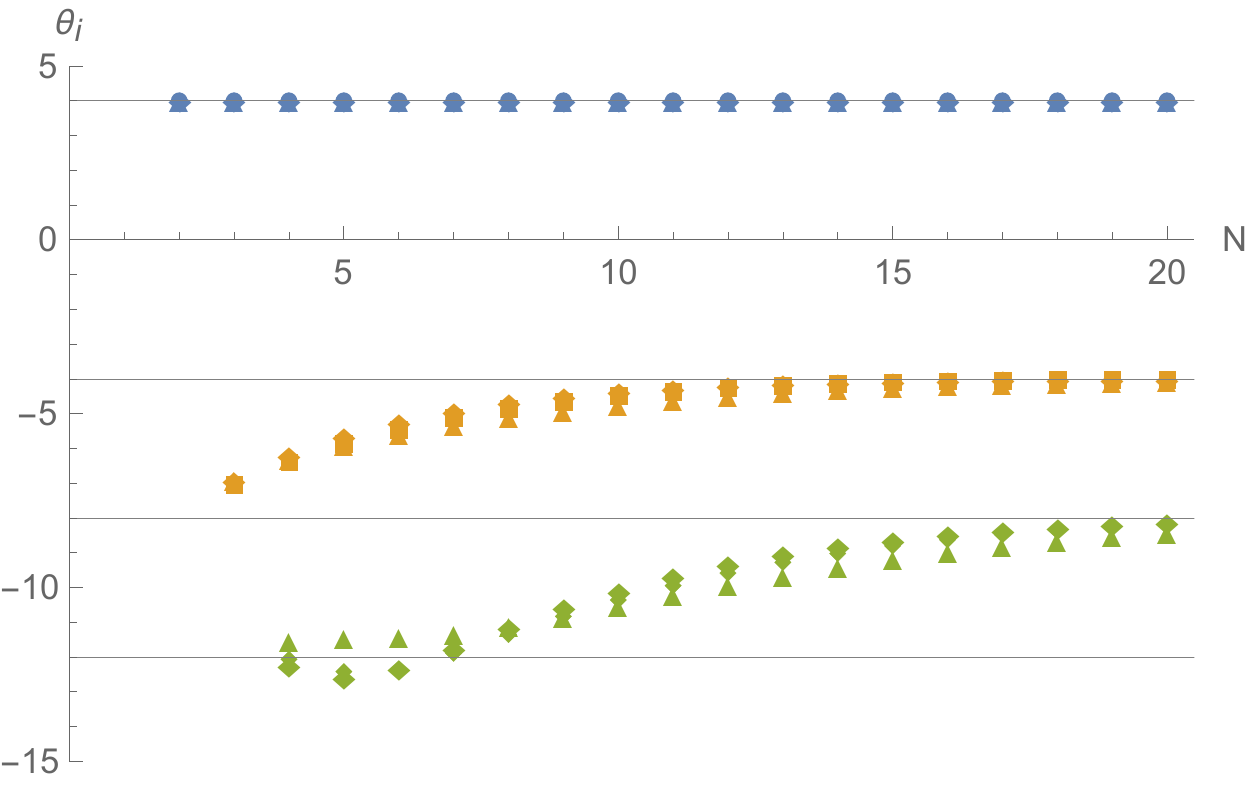} \\
	\caption{\label{fig.3} Stability coefficients of the NGFP$_1$ arising in the SSA applied to scalar fields (circles), gauge fields (diamonds), and the $O(N_\varphi)$-model with $N_\varphi = 2$ (triangles) up to order $N=20$. The coefficients are ordered by magnitude $\theta_i > \theta_{i+1}$ with $\theta_1$, $\theta_2$, and $\theta_3$ given in the blue, orange, and green series, respectively. The gray lines indicate the asymptotic values of the series.}
\end{figure}

Fig.\ \ref{fig.etas} indicates that neglecting the contribution of the anomalous dimension actually constitutes a good approximation, at least for sufficiently large values $N$. Substituting $\eta_i = 0$ into the generating functionals simplifies the beta functions significantly such that one can extend the number of couplings which can be tracked with a reasonable numerical effort. In practice, we then follow the simplified system up to $N=20$.  While there is no conceptual problem in going beyond this setting, the length of the expressions makes this analysis computationally expensive, so we limit ourselves to this setting. 

The stability coefficients $\theta_i$ obtained in this setting are shown in Fig.\ \ref{fig.3}. The UV-relevant stability coefficient has $\theta_1 = 4$, independently of $N$. For sufficiently large values of $N$, the $N$-dependence of $\theta_2$ and $\theta_3$ is well-captured by the asymptotics
\be\label{fitfunction}
\theta_i = a_i + \frac{b_i}{N^4} \, . 
\ee
Here the exponent is fixed based on the observed fall-off properties of the data points. In order to find the fit-parameters $a_i$ and $b_i$ we then go into the ``tail'' of the curve where the stability coefficients already show convergence. Practically, we chose to incorporate the values for $\theta_i$ obtained for $N \ge 10$ and checked that there is no qualitative difference if one includes or excludes one or two additional points. The best-fit values obtained in this way are listed in Table \ref{tab.fitparameters}.
\begin{table}[h!]
	\centering
	\begin{tabular}{cccc}
		\hline
		& & \quad $a_i$ \quad & \quad $b_i$ \quad \\
		\hline
		\multirow{3}{*}{$\theta_2$}& scalar &  $-3.99$  &  $-5204$  \\
		& $O(2)$ & $-4.04$ & $-7477$ \\
		&gauge field & $-3.98$ & $-3975$ \\
		\hline
		\multirow{3}{*}{$\theta_3$}& scalar & $-8.25$ & $-23606$ \\
		& $O(2)$ & $-8.57$ & $-22672$ \\
		&gauge field & $-8.14$ & $-21584$ \\
		\hline
	\end{tabular}
	\caption{\label{tab.fitparameters} Parameters derived from fitting the stability coefficients $\theta_2$ and $\theta_3$ obtained for $N \ge 10$ to the profile \eqref{fitfunction}.}
\end{table}
The values $a_i$ give an estimate for the ``best'' value of the stability coefficient in the limit $N\rightarrow \infty$. We observe that the convergence of $\theta_2$ (and $\theta_3$) is already very good while the stabilization of $\theta_n$, $n \ge 4$ requires truncations with $N > 20$. 

Heuristically, the appearance of a relevant eigendirection can be understood as follows. Switching off the anomalous dimension, $\eta_i = 0$, and setting $u_n = 0, n \ge 3$, the beta function governing the scale-dependence of $u_2$ has the form
\be\label{beta2}
\beta_{u_2} = u_2 \left(a + b u_2 \right) \, , 
\ee
with positive coefficients $a,b$ whose precise value depends on the system under consideration. Thus $\beta_{u_2}$ is a quadratic polynomial with roots at $u^*_2 = 0$ (GFP) and $u_2^* = - \frac{a}{b} < 0$ (NGFP$_1$). The stability coefficients are given by minus the slope of the polynomial at these points. It is then straightforward to verify that they must be equal in magnitude and opposite in sign $\theta_1^{\rm GFP} = - \theta_1^{{\rm NGFP}_1}$. Thus in the specific situation \eqref{beta2}, $\theta_1^{{\rm NGFP}_1}$ is fixed by the mass-dimension of the coupling and has the opposite stability properties encountered at the GFP. 

At this stage a remark on the convergence of the expansion \eqref{polynomial} is in order. Within the present analysis, we constructed the series up to $N=20$ and investigated the ratio $u_{n+1}/u_n$ to estimate the radius of convergence. Within this analysis, we did not find evidence that the ratio converges to a finite limit. Since all $u_n$ are negative there are also no cancellations between neighboring terms which could improve the radius of convergence. Hence it is likely, that $\tilde{f}_*(X)$ linked to the solutions NGFP$_1$ is actually an asymptotic series with zero radius of convergence.

\section{Testing for inessential operators}
\label{sect.redundancy}
Given the peculiar structure of the eigenvalue spectrum found for the stability matrix ${\bf B}$ shown in Fig.\ \ref{fig.3}, it is important to clarify whether all deformations of the fixed point are physical. Following \cite{Wegner_1974,Hawking:1979ig,Dietz:2013sba}, this question can be addressed by searching for so-called inessential (also called redundant) operators $O_r$. Formally, an inessential operator is
\begin{enumerate}
	\item an eigenoperator of the stability matrix,
	\item equivalent to an infinitesimal change of a field variable \, . 
\end{enumerate}
The second property implies that $O_r$ takes the form
\be\label{eq.redundantop}
O_r = \int d^dx \frac{\delta \Gamma_*}{\delta \phi} \, F[x^\mu,\phi] \, , 
\ee
where $\phi \mapsto \phi + \epsilon F[x,\phi]$ with $F[x,\phi]$ a function of position $x^\mu$ and a (not necessarily local) functional of $\phi$. For the $f(X)$-truncations considered in this work, the field redefinitions of interest have the form $F[\phi] = \frac{1}{2} \phi \tilde{F}(X)$ where the factor $1/2$ is extracted for convenience. Taking into account that $X$ is constant within our approximation, eq.\ \eqref{eq.redundantop} evaluates to
\be\label{eq.redundantopev}
O_r(X) = \int d^dx \, X \, \frac{\p f(X)}{\p X} \, \tilde{F}(X) 
\ee
Thus, we can always find a function $\tilde{F}(X)$ solving this relation, \emph{unless $ \frac{\p f(X)}{\p X}$ is not invertible}. While the potential lack of convergence in the polynomial expansions constructed in the present work makes it impossible to draw any final conclusion about zeros of $\frac{\p f(X)}{\p X}$, we can still make the following tentative observations: firstly, at $X=0$ the normalization of the kinetic term ensures that $\frac{\p f(X)}{\p X} > 0$. Secondly, all coefficients appearing in the polynomial expansion of $\tilde{f}(X)$ are negative, indicating that at a sufficiently large value $X$ we must have $\frac{\p f(X)}{\p X} < 0$. Continuity of $f(X)$ would then entail that there must be a zero of the first derivative. Based on this argument, one expects that the deformations constructed at the NGFP$_1$ do not correspond to inessential operators. 

Note that this argument does not apply at the GFP though: in this case $\frac{\p f(X)}{\p X} = 1$ and deformations of $f(X)$-type are inessential by the argument \eqref{eq.redundantopev}. Thus we observe that the definition of an inessential operator \emph{actually depends on the structure of the fixed point under consideration}. This has profound consequences for the essential scheme \cite{Baldazzi:2021ydj,Baldazzi:2021orb} where the decision about inessential operators is made before performing the RG computation.
\section{Conclusions and outlook}
\label{sect.implications}
Motivated by recent results on renormalization group fixed points for scalar-tensor theories \cite{deBrito:2021pyi,Steinwachs:2021jft,Laporte:2021kyp,Knorr:2022ilz}, we used the Wetterich equation for the effective average action $\Gamma_k$ to study the renormalization group (RG) flow of $O(N_\varphi)$ scalar field theory and Abelian gauge fields employing a novel shift-symmetric approximation (SSA). The characteristic feature of this projection is that $\Gamma_k$ is approximated by a function of the kinetic term $X$. In contrast to the local potential approximation (LPA) typically employed when studying scalar field theory, this ansatz is compatible with the global symmetries of the kinetic term. This makes the approximation predestined for exploring the structure of matter self-interactions induced by gravitational fluctuations.

Studying the fixed point structure in a polynomial expansion, we find evidence for \emph{a new universality class realized in all matter systems under consideration}. This class is distinguished by its very special stability coefficients $\theta_n$: the current investigation identifies one UV-relevant eigendirection $\theta_1 = 4$ whose stability coefficient has the opposite sign from the one encountered at the free field fixed point (GFP). All other stability coefficients determined in our work are in agreement with the spectrum of irrelevant deformations at the GFP. Increasing the number of operators tracked in the computation the NGFP shifts towards the GFP. The specific values of the coupling constants suggest that the function $\tilde{f}_k$ encoding the momentum-dependent interactions actually constitutes an asymptotic series when expanded at $X=0$. While our letter reports the analysis in $d=4$ spacetime dimensions, we also verified that the NGFP$_1$ and its key properties also persist in $d=3$.

Clearly, having evidence for a novel universality class for matter systems is interesting in its own right. In addition, our result has far-reaching consequences for gravity-matter systems investigated within the gravitational asymptotic safety program. At the technical level, the new universality class may constitute the entry point for studying questions related to asymptotic safety based on conventional conformal field theory techniques like operator product expansions.\footnote{For recent works focusing on the connection between conformal field theory and functional renormalization group methods, see \cite{Pagani:2017tdr,Codello:2017hhh,Codello:2017epp,Codello:2018nbe,Vacca:2019rsh,Pagani:2020ejb,Rose:2021zdk}.} Moreover, the universality class studied in our work plays a crucial role in bounding the values of effective couplings compatible with asymptotic safety \cite{Knorr:2022ilz}. Moreover, the peculiar stability properties of the NGFP imply that it acts as a very efficient funnel: any RG trajectory coming in its vicinity is captured and emitted along a very specific direction when continuing towards the infrared. This funnel-mechanism has profound consequences for the predictive power of asymptotic safety as it efficiently wipes out information about the high-energy origin of the flow. We hope to come back to these intriguing points in the near future.

\section*{Acknowledgements}
\noindent
We thank T.\ Budd for interesting discussions and B.\ Knorr for insightful comments on the manuscript. A.D.P. acknowledges CNPq under the grant PQ-2 (309781/2019-1), FAPERJ under the “Jovem Cientista do Nosso Estado” program (E26/202.800/2019), and NWO under the VENI Grant (VI.Veni.192.109) for financial support. The work by C.L.\ is supported by the scholarship Becas Chile ANID-PCHA/2020-72210073. 

\begin{appendix}
\section{Technical Annex}
\label{app.a}
A characteristic feature of the SSA is the occurrence of loop-momenta which are contracted with derivatives of the background field. Within the momentum integral, such terms can be reduced to integrals over the squared momentum, replacing 
\be\label{eq.prepgen}
\begin{split}
& p^{\mu_1} p^{\nu_1}  \cdots p^{\mu_n} p^{\nu_n} \\ & \mapsto \frac{(p^2)^n}{d(d+2)\cdots(d+2n-2)} \underbrace{\left[\delta_{\mu_1\nu_1} \cdots \delta_{\mu_n \nu_n} + \cdots \right]}_{ (2n-1)!! \, \text{permutations}} \, .
\end{split}
\ee
Here we use the $\mapsto$ to indicate that this identity holds under the momentum-integral only. The identities \eqref{eq.prepgen} generalize the relations for $n=1$ and $n=2$ \cite{Peskin:1995ev}
\be
\begin{split}
p^\mu p^\nu \mapsto & \, \frac{p^2}{d} \delta^{\mu\nu} \, , \\
p^\mu p^\nu p^\rho p^\sigma \mapsto & \, \frac{p^4}{d(d+2)}  \left[\delta^{\mu\nu} \delta^{\rho\sigma} + \delta^{\mu\rho} \delta^{\nu\sigma} + \delta^{\mu\sigma} \delta^{\rho\nu}\right] \, .
\end{split}
\ee

When performing resummations of an infinite series, it is convenient to express the combinatorical prefactors associated with \eqref{eq.prepgen} in terms of Gamma-functions
\be\label{eq.ids}
\begin{split}
\frac{1}{d(d+2)\cdots(d+2n-2)} = & \frac{\Gamma(d/2)}{2^n \, \Gamma(d/2+n)} \, , \\
(2n-1)!! = & \frac{\Gamma(2n)}{2^{n-1} \Gamma(n)} \, . 
\end{split}
\ee
In the context of scalar fields, it is then straightforward to derive the replacement rule
\be\label{eq.ssamaster}
\begin{split}
 \left( p^{\mu} p^{\nu} (\p_{\mu} \phi) (\p_{\nu} \phi) \right)^n  & \, \mapsto 2 \, \frac{ \Gamma(d/2) }{\Gamma(d/2+n)} \, \frac{\Gamma(2n)}{ \Gamma(n)} \, \left( \frac{1}{2Z_k} \, p^2 \, X \right)^n \, . 
\end{split}
\ee
 The derivation in the context of Abelian gauge fields utilizes a similar identity, substituting
\be\label{eq.ssagauge}
\left( p^{\mu} p^{\nu} F_\mu{}^\alpha F_{\nu\alpha}  \right)^n \mapsto 
\frac{ \Gamma(d/2) }{\Gamma(d/2+n)} \left( \frac{2}{Z_k} \, p^2 \, X \right)^n \, . 
\ee 
Note that the combinatorics in the case of scalar fields and gauge fields is actually quite different. In the former case all $(2n-1)!!$ permutations appearing in \eqref{eq.prepgen} contribute equally. In the latter case, it is only the first term in \eqref{eq.prepgen} which gives rise to the proper index contraction. All other index combinations give strings of $F$'s build from more than two field strength tensors. Since these are not tracked in the present work, they are not included in the projection \eqref{eq.ssagauge}.
\end{appendix}

\bibliography{general_bib}
\end{document}